\documentclass[conference,letterpaper]{IEEEtran}
\IEEEoverridecommandlockouts

\usepackage[final]{microtype}
\usepackage{amsmath,amssymb,amsfonts}
\usepackage{algorithm}
\usepackage{algpseudocode}
\usepackage{booktabs}
\usepackage{graphicx}
\usepackage{makecell}
\usepackage{tabularx}
\usepackage{array}
\usepackage{cite}
\usepackage{balance}
\usepackage{tikz}
\usetikzlibrary{arrows.meta,positioning}
\PassOptionsToPackage{hyphens}{url}
\usepackage[hidelinks]{hyperref}
\def\BibTeX{{\rm B\kern-.05em{\sc i\kern-.025em b}\kern-.08em
    T\kern-.1667em\lower.7ex\hbox{E}\kern-.125emX}}

\begin{document}

\makeatletter
\def\bstctlcite#1{\@bsphack
  \@for\@citeb:=#1\do{\edef\@citeb{\expandafter\@firstofone\@citeb}%
    \if@filesw\immediate\write\@auxout{\string\citation{\@citeb}}\fi}%
  \@esphack}
\makeatother
\bstctlcite{IEEEexample:BSTcontrol}

\title{PRB Starvation at Slot Time via\\
Rank Indicator Inflation: Attacking\\
Proportional Fair Memory in the Scheduler
\thanks{Research supported by the NVIDIA Academic Grant Program using DGX Spark, Sionna, and Sionna-RT. It is also supported in part by the State of South Carolina through the Battelle Savannah River Alliance Workforce Development Program and the National Science Foundation under Grants CNS-2202972, CNS-2318726, and CNS-2232048.}}

\author{%
\IEEEauthorblockN{Ryan Barker\IEEEauthorrefmark{1},
Julia Boone\IEEEauthorrefmark{1},
Tolunay Seyfi\IEEEauthorrefmark{1},
Alireza Ebrahimi\IEEEauthorrefmark{1},\\
Ethan Farquhar\IEEEauthorrefmark{2},
Brian Davis\IEEEauthorrefmark{2},
Joseph Boccuzzi\IEEEauthorrefmark{3},
Fatemeh Afghah\IEEEauthorrefmark{1}}
\IEEEauthorblockA{\IEEEauthorrefmark{1}\textit{Holcombe Department of Electrical and Computer Engineering}, \textit{Clemson University}\\
Clemson, SC 29634, USA\\
rcbarke@clemson.edu, jcboone@clemson.edu, tseyfi@clemson.edu, alireze@clemson.edu, fafghah@clemson.edu}
\IEEEauthorblockA{\IEEEauthorrefmark{2}\textit{Savannah River National Laboratory}, Jackson, SC 29831, USA\\
ethan.farquhar@srnl.gov, brian.davis@srnl.gov}
\IEEEauthorblockA{\IEEEauthorrefmark{3}\textit{NVIDIA Corporation}, Santa Clara, CA 95051, USA\\
jboccuzzi@nvidia.com}
}

\maketitle

\begin{abstract}
Proportional fair (PF) scheduling in fifth-generation (5G) New Radio (NR) ranks user equipment (UE) by instantaneous rate over an exponentially weighted moving average (EWMA) of delivered service. We identify a served-but-undelivered denial-of-service (DoS) attack in which rank indicator (RI) inflation induces persistent physical resource block (PRB) starvation. Unsupported spatial layers cause first-attempt delivery failure, keeping the attacker's service EWMA low; the evaluated PF allocator repeatedly grants it the full PRB pool, denying honest UEs transmission opportunities. We formalize this recurrence and evaluate it in 3,300 paired, 200 slot multi-user shards spanning rank 1 and rank 2 Sionna Ray Tracing geometries, three link adaptation controllers, PF and round robin (RR), four target identities, and five seeds. On the rank 2 geometry, selective RI inflation yields 94--95\% target capture, reduces PRB Jain fairness from 0.997 to 0.28, and produces mean honest zero-grant runs of 41--56\,ms; RR preserves equal allocation. Exact offline replay and a Monte Carlo recurrence isolate the delivered-service denominator as the causal path. Cross-field and controller ablations show a weaker, controller-dependent channel-quality trigger, while activity-gated EWMA defenses remain structurally blind. The results motivate RI-consistency checks before feedback acquires scheduling authority and slot-time PRB evidence inside the O-RAN Distributed Unit.
\end{abstract}

\begin{IEEEkeywords}
Open Radio Access Network, 5G New Radio, Proportional Fair scheduling, PRB starvation, slot time telemetry, scheduler native metrics, Sionna.
\end{IEEEkeywords}

\section{Introduction}
\label{sec:intro}
Mission-critical 5G/6G services, including industrial control and Ultra Reliable Low Latency Communications (URLLC), depend on continuous service at scheduler timescale \cite{TS22261,9162761}. A cell can nevertheless maintain acceptable second-scale throughput or fairness while an honest user equipment (UE) receives no transmission grant for tens of milliseconds. In an Open Radio Access Network (O-RAN), interoperable telemetry and control remain largely above the sub-10 ms decisions made inside the O-RAN Distributed Unit (O-DU) \cite{ORAN_E2SM_KPM,GarciaSaavedra2021ORAN}. We therefore ask whether falsified channel state information (CSI) can cause a scheduler to repeatedly reward a UE that consumes resources but delivers no service, and whether the resulting denial-of-service condition can be isolated at slot time.

The attack surface is proportional fair (PF) memory. PF divides an instantaneous-rate estimate by an exponentially weighted moving average (EWMA) of delivered service. Earlier starvation studies reached this edge through online traffic shaping: an idle-then-bursty UE lets its average forget prior service and resumes with elevated priority \cite{yang2004spfs,bali2008paq}. We expose a second path that requires no payload transmission and originates in CSI authority. Downlink link adaptation uses the rank indicator (RI) to select spatial layers \cite{TS38214,pedersen2009downlink}. In the evaluated model, unsupported spatial layers cause failed delivery that consumes physical resource blocks (PRBs) without raising the target's service average. PF repeatedly favors the target, denying honest UEs transmission opportunities: denial of service (DoS) and network brown-out conditions through scheduler-amplified PRB starvation.

The execution regime is part of this claim. AtlasRAN shows that host-operating-system radio emulation can support protocol tests and controlled functional ablations, but real-time-factor (RTF) dilation, input/output backpressure, and instrumentation loss can invalidate wall-clock deadline or deployment-capacity inference \cite{atlasran2026}. We therefore use modeled slot progression in a paired Sionna physical layer and ray tracing harness. The claim is deliberately causal and policy-specific: it concerns scheduler behavior under the stated PF recurrence, not over-the-air attack realization, fronthaul timing, or deployed cell capacity.

This paper makes four contributions. First, it formalizes a served-but-undelivered PF edge and derives a closed-form starvation horizon, extending classical idle/burst analysis to a continuously scheduled UE. Second, it performs a 3,300 shard paired ablation across two multi-user ray tracing geometries, three link  adaptation controllers, PF and round robin (RR), four target identities, five seeds, and RI, channelquality, Hybrid Automatic Repeat Request (HARQ), and precoder controls. Under selective rank 2 RI inflation, PF assigns 94--95\% of downlink slots to the target and reduces PRB Jain fairness from 0.997 to 0.28. Third, exact schedule replay and a target-wise Monte Carlo recurrence verify that the channel-grounded numerator is unchanged and that delivery failure reaches allocation through the EWMA denominator. Fourth, the analysis shows why activity-gated EWMA defenses do not engage and identifies the required defense boundary: RI-consistency verification before allocation, paired with a slot-time PRB ledger inside the O-DU.

\section{PF Memory and the Feedback Attack Surface}
\label{sec:background}
PF remains common in cellular scheduling because it balances multi-user diversity and long-term service by normalizing instantaneous opportunity with a smoothed service history \cite{3gpp_pf}. The Third Generation Partnership Project (3GPP) standardizes procedures and measurements but does not prescribe a single, universal base-station scheduler \cite{TS38214}. Silence-aware proportional scheduling and the starvation-resistant proportional allocation with quantiles (PAQ) method reduce burst-entry priority by freezing or shadowing the average-rate state during inactivity \cite{yang2004spfs,bali2008paq}. These methods address an idle attacker. They do not address a UE that remains backlogged, receives every grant, and repeatedly delivers zero bits.

CSI integrity work shows that syntactically valid feedback can still be semantically false. RI tampering can induce spatial misconfiguration \cite{timilsina2025csitampering}, compressed-feedback poisoning can disrupt MU-MIMO beamforming \cite{meneghello2024whack}, and low-power signaling can alter receiver or scheduler behavior without broad-band jamming \cite{ludant2021sigunder}. We do not implement these waveform paths. We inject their scheduler-visible RI or channel-quality consequences before the same-slot decision and measure the allocation response.

The observability problem is architectural. Near-real-time RAN Intelligent Controller (RIC) loops operate above the O-DU scheduler cadence, whereas distributed applications (dApps) colocated with the O-DU are proposed for faster inference and control \cite{ORAN-RR-2024-10-dApps}. A one-second KPI can average over twenty 50 ms deprivation episodes. A slot-time PRB ledger retains their continuity and exposes whether a fairness loss came from channel opportunity, scheduling policy, or delivery failure.

\section{Threat Model and Slot-Time Analysis}
\label{sec:model}
\subsection{System and Adversary}
\begin{figure}[!tb]
\centering
\includegraphics[width=\columnwidth]{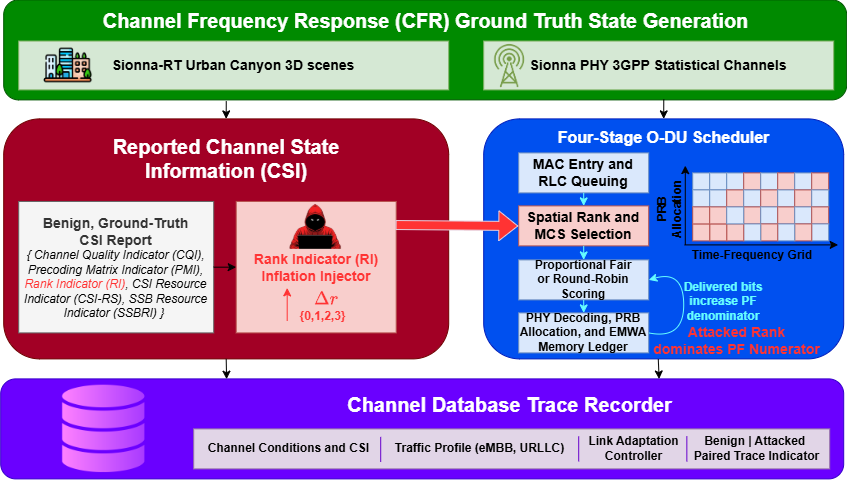}
\caption{Paired effect-injection harness. Sionna PHY and ray tracing sources generate channel-grounded state. Honest and attacked legs share channel, traffic, controller, target, and random state; the injector modifies only the selected report field. The allocator and delivered-bit ledger close the PF-memory loop. Multi-user results use modeled-time ray-tracing geometries, not an over-the-air attack implementation.}
\label{fig:architecture}
\end{figure}

The harness in Fig.~\ref{fig:architecture} serves four UEs in a single cell on a virtual 273 PRB reference carrier with 30 kHz subcarrier spacing and numerology $\mu=1$, which gives a 0.5 ms slot. A 40 slot time-division duplex (TDD) period contains 28 downlink slots. In each downlink slot, the allocator grants the full PRB pool to one UE. The primary attack is selective: three UEs report honestly and one target modifies only its own RI before the same-slot scheduling decision. The target observes its honest report and the codebook cap $r_{\max}$, but not the other UEs, allocator state, random seed, or future channel.

The primary channel supports rank $r_i=2$ for each UE while the codebook allows up to four scheduled layers. The target reports
\begin{equation}
\widehat r_i[t]=\min\!\left(r_{\max},r_i[t]+\Delta_r\right),
\qquad \Delta_r\in\{0,1,2,3\}.
\label{eq:ri}
\end{equation}
The rank 1 ray tracing control uses one transmit dimension and four receive dimensions per UE, so its codebook cap is one and RI inflation is clipped before scheduling. We retain this lockstep control to separate an ineffective report from the rank 2 geometry, which has spatial headroom to over-claim.

\subsection{PF Memory and the Served-but-Undelivered Edge}
Algorithm~\ref{alg:capture} summarizes the capture loop. The evaluated PF allocator scores UE $i$ at slot $t$ by
\begin{equation}
 s_i[t]=\frac{\mathrm{SE}_i[t]}{\overline R_i[t]+\varepsilon},\qquad
 \overline R_i[t+1]=(1-\alpha)\overline R_i[t]+\alpha q_i[t]\phi_i[t],
\label{eq:pf}
\end{equation}
where $\mathrm{SE}_i[t]=\log_2(1+\gamma_i[t])$ uses channel-grounded signal-to-interference-plus-noise ratio (SINR) $\gamma_i[t]$, $\phi_i[t]$ is the granted transport-block bit count, and $q_i[t]\in\{0,1\}$ is first-attempt delivery. The guard $\varepsilon=1$ bounds the first-slot score. The update factor is $\alpha=0.05$, giving time constant $\tau=1/\alpha=20$ modeled slots, or 10\,ms. Because the numerator never consumes the falsified report, any allocation change enters through delivered service in the denominator.

If a UE receives no delivered bits for $T$ updates, whether idle or unsuccessfully served,
\begin{equation}
\overline R_i[t+T]=(1-\alpha)^T\overline R_i[t]
\approx e^{-T/\tau}\overline R_i[t].
\label{eq:decay}
\end{equation}
For $\Delta_r\ge 1$ in the evaluated rank 2 geometry, unsupported layers dominate the modeled block error probability. The measured target therefore has $q_i[t]=0$ on each served slot, so $\overline R_i[t]\rightarrow 0$ while it consumes the full grant and $s_i[t]\rightarrow \mathrm{SE}_i[t]/\varepsilon$. This is the same geometric memory edge as an idle interval, but it is persistent because there is no traffic burst completion or backlog drain to end it.

After an honest UE $j$ wins once and delivers $\phi_j$ bits, its EWMA is approximately $\alpha\phi_j$. Following $n$ unserved updates, it can strictly outrank the pinned target only if $\mathrm{SE}_j>\mathrm{SE}_i$ and
\begin{equation}
 n_{\mathrm{starve}}(j)=
 \left\lceil
 \frac{\ln\!\left(\alpha\phi_j/[\varepsilon(\mathrm{SE}_j/\mathrm{SE}_i-1)]\right)}
 {-\ln(1-\alpha)}
 \right\rceil_{+},
\label{eq:horizon}
\end{equation}
where $\lceil x\rceil_{+}=\max(0,\lceil x\rceil)$. If $\mathrm{SE}_j\le\mathrm{SE}_i$, the strict horizon is unbounded absent a target decode, a tie-break, or another policy intervention. The measured bit counts place this horizon between 150 and 219 modeled updates across target identities, explaining why recovery depends on rare target outcomes rather than PF's nominal fairness memory.

\begin{algorithm}[t]
\caption{rank inflation capture at 0.5 ms slot time}
\label{alg:capture}
\begin{algorithmic}[1]
\Require supported rank $r_i$, cap $r_{\max}$, inflation $\Delta_r\ge1$
\State Target reports $\widehat r_i\gets\min(r_{\max},r_i+\Delta_r)$; other CSI is unchanged.
\State PF selects $\arg\max_k \mathrm{SE}_k/(\overline R_k+\varepsilon)$; the winner receives all 273 PRBs.
\State Modeled delivery fails for the target; set $q_i=0$ and update $\overline R_i$ using~\eqref{eq:pf}.\State The target score approaches $\mathrm{SE}_i/\varepsilon$; honest averages must decay before readmission.
\State Record per-slot winner, PRBs, decode, and EWMA state for replay and run-length metrics.
\end{algorithmic}
\end{algorithm}

\subsection{Scheduler-Native Metrics}
Winner-take-all allocation makes instantaneous four UE Jain fairness equal to $1/4$ in every downlink slot, independent of policy. We therefore sum each UE's PRBs over a window $W$ and compute
\begin{equation}
G_i(W)=\sum_{t\in W}\mathrm{PRB}_i[t],\qquad
J_W=\frac{(\sum_iG_i(W))^2}{4\sum_iG_i^2(W)}.
\label{eq:jain}
\end{equation}
We report target win share, $J_{200}$ over the full shard, the minimum $J_{20}$ over 20 consecutive downlink decisions, and the longest contiguous zero-grant run for each honest UE. Each shard has 200 modeled slots and 140 downlink slots, so run lengths are censored at 70\,ms. Throughput Jain fairness is computed separately from delivered bits to show when aggregate service equality masks allocation capture \cite{jain1984}.

\section{Experimental Method}
\label{sec:method}
The analyzed sweep separates single UE physical layer (PHY) controls from multi-user Sionna Ray Tracing (RT) surfaces \cite{hoydis2023sionnaopensourcelibrarynextgeneration,hoydis2023sionnartdifferentiableray}. The PHY controls are additive white Gaussian noise (AWGN) with a $1\times1$ rank 1 link, Third Generation Partnership Project (3GPP) Tapped Delay Line C and D (TDL-C/D) non-line-of-sight (nLoS) and line-of-sight (LoS) channels with $2\times2$ rank 2 links, and a Clustered Delay Line A (CDL-A) mixed channel with a $2\times4$ rank 2 link \cite{3gpp38901}. They isolate controller and feedback response but cannot establish multi-user PRB starvation. The allocation claims use two stationary urban-canyon RT geometries with four UEs: a $1\times4$ per-UE rank 1 control and a $2\times4$ per-UE rank 2 spatial surface, with mean channel-grounded SINR from 9.6 to 11.7\,dB.

Every channel is evaluated with outer loop link adaptation (OLLA), self-adaptive link adaptation (SALAD), or SALAD with reported channel quality information in its modulation and coding scheme (MCS) path (SALAD-CQI) \cite{pedersen2009downlink,wiesmayr2026salad}. A type I codebook selector chooses spatial rank. Multi-user RT cells then apply PF or RR to a virtual 273 PRB, 100 MHz, 30 kHz reference carrier. The Sionna grid has 14 orthogonal frequency-division multiplexing (OFDM) symbols and 64 tones and yields scalar effective SINR. PRB accounting is frequency-flat and does not alter SINR; inter-user interference is not modeled.

The multi-user audit contains 3,300 paired 200 slot shards. The rank 2 RT surface evaluates selective RI inflation at $\Delta_r=0$--3 and PMI randomization off/on. The rank 1 RT surface evaluates selective CQI/SINR bias at 0, 3, 6, 9, 12, 15--21 in 1 dB steps, 24, 27, and 30\,dB; lockstep HARQ flips at $p\in\{0,.05,.10,.15,.25,.50\}$; and clipped lockstep RI controls. Matched null legs are retained. Every cell spans three link adapters, PF/RR, and five seeds; selective cells also sweep target UE $0$--$3$. Selective rank 2 RI is the primary hypothesis, CQI/SINR bias tests cross-field generality, and HARQ/PMI are specificity controls that are not pooled with the headline result.

\begin{figure*}[t]
\centering
\includegraphics[width=0.70\textwidth]{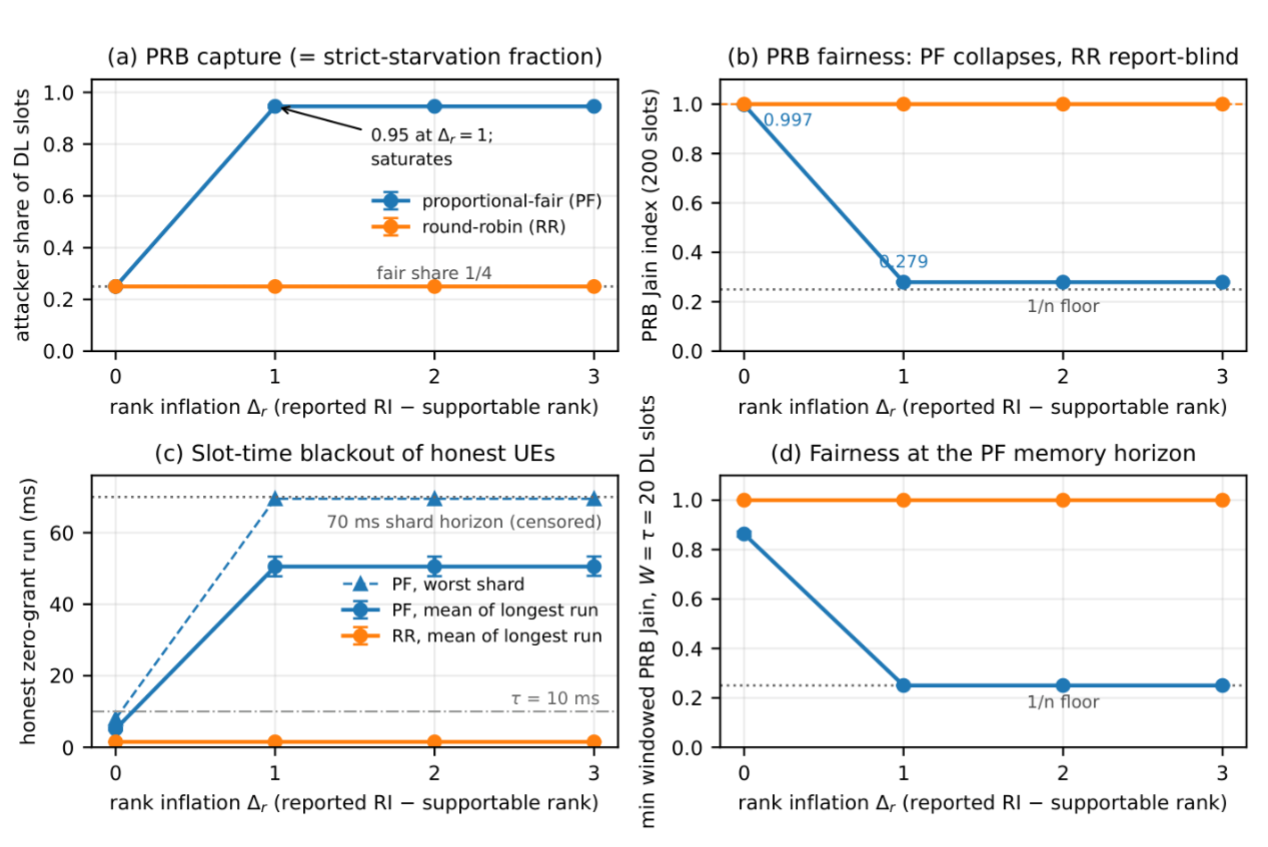}
\caption{Selective RI inflation on the rank 2 urban-canyon geometry. Each point pools 60 shards across four target identities, three link adapters, and five seeds; error bars are 95\% shard-bootstrap intervals. Target capture and both fairness metrics saturate at $\Delta_r=1$. PF reaches the four UE fairness floor and produces long honest-user blackouts, while RR remains report-blind in its PRB rotation.}
\label{fig:results}
\end{figure*}

Honest and attacked legs share channel, traffic, mobility, controller, target, and random state; the injector changes only the selected report field. Recorded fields include channel-grounded SINR, spectral efficiency, granted bits, first-attempt decode, winner, PRB count, scheduled rank, and PF state. Offline replay recomputes every PF decision, and a Monte Carlo recurrence draws Bernoulli decodes at each UE's measured rate. The complete 200 slot shard is the replication unit. Fig.~\ref{fig:results} reports 95\% percentile-bootstrap intervals over 60 shards per point, while Table~\ref{tab:modelcheck} retains target identities separately. Slots are not treated as independent samples.

\section{Results and Discussion}
\label{sec:results}

\subsection{Capture, Fairness, and Saturation}
Fig.~\ref{fig:results}(a) shows the direct scheduling effect. Under honest PF, each target wins approximately one quarter of downlink slots. At $\Delta_r=1$, pooled target share rises to 0.946 and remains identical at strengths two and three. Across target identities, the measured range is 0.940--0.953, while the controller means are 0.955 for OLLA, 0.923 for SALAD, and 0.960 for SALAD-CQI. One unsupported layer is therefore sufficient to zero first-attempt delivery and saturate the PF denominator path across all three controllers. Higher strengths change scheduled rank from three to four but leave scheduling, Jain, and run-length metrics unchanged.

Allocation fairness collapses much more sharply than throughput fairness (Table~\ref{tab:summary}). PF PRB $J_{200}$ falls from 0.997 to a pooled mean of 0.279, and the minimum $J_{20}$ reaches its winner-take-all floor of 0.25. The pooled mean longest honest zero-grant run is 50.5\,ms; target-specific means span 41.1--55.5\,ms, with observed maxima from 44.5 to 69.5\,ms. Between 68.9\% and 83.5\% of honest runs exceed the 10-ms PF time constant. RR holds target share at 0.25, PRB $J_{200}$ at 1.0, and longest runs at three downlink slots, or 1.5\,ms. Under RR, the target's zero decode rate lowers delivered-throughput Jain fairness to about 0.748, but the target harms itself rather than capturing honest PRBs.

Throughput fairness masks the allocation mechanism. PF throughput $J_{200}=0.715$--0.729 is near RR's 0.748 because either policy contains one zero-delivery UE. The PRB ledger separates them: RR wastes the target's fixed quarter, whereas PF redirects about 70\% of additional opportunities away from honest UEs. Delivered throughput alone cannot distinguish self-inflicted failure from scheduler-amplified starvation.

\begin{table}[!tb]
\caption{Selective rank 2 RI Inflation. Values Are Ranges Across Four Target Identities; Each Cell Contains 15 Shards. Run Length Is Mean (Maximum) in Milliseconds.}
\label{tab:summary}
\centering
\footnotesize
\setlength{\tabcolsep}{1.7pt}
\begin{tabular}{@{}lccccccc@{}}
\toprule
Policy & $\Delta_r$ & \makecell{Target\\win} & \makecell{PRB\\$J_{200}$} & \makecell{Tput.\\$J_{200}$} & \makecell{Min.\\$J_{20}$} & \makecell{Run\\ms} & Decode \\
\midrule
PF & 0 & .241--.259 & .997 & .999 & .863 & \makecell{4.7--5.3\\(8.0)} & .88--.89 \\
PF & 1--3 & .940--.953 & .275--.283 & .715--.729 & .250 & \makecell{41.1--55.5\\(69.5)} & .00 \\
RR & 0 & .250 & 1.000 & .997 & 1.000 & \makecell{1.5\\(1.5)} & .88--.89 \\
RR & 1--3 & .250 & 1.000 & .748--.749 & 1.000 & \makecell{1.5\\(1.5)} & .00 \\
\bottomrule
\end{tabular}
\end{table}

RI inflation is inert on the rank 1 control at every strength. Its scheduled rank remains one, target decode remains approximately 0.92, PF PRB $J_{200}$ remains 0.995, and win shares remain near 0.24, 0.24, 0.26, and 0.26. The exploit therefore requires spatial headroom between the true channel support and the codebook rank cap. The other feedback controls do not reproduce the collapse: nonzero lockstep HARQ flipping retains mean PF PRB $J_{200}=0.994$, while selective PMI randomization on the rank 2 surface yields mean target share 0.315 and PRB $J_{200}=0.973$. These specificity checks and the rank 1 null are not pooled with the selective RI claim.

\subsection{Mechanism Verification}
Replaying Eq.~\eqref{eq:pf} on each downlink decision while advancing the EWMA on every modeled slot reproduces 100.0\% of recorded PF winners in every selective rank 2 cell. Advancing the EWMA only on downlink slots reduces match to 0.981--0.983. The schedule is therefore fully explained by the stated PF recurrence and its all-slot update cadence. Because $\mathrm{SE}_i$ is channel-grounded, the falsified RI reaches allocation only through $q_i\phi_i$ and the denominator $\overline R_i$.

The Monte Carlo recurrence predicts target capture of 0.943--0.995 against 0.940--0.953 measured, with per-cell win-share error at most 0.022. Predicted longest runs are 115.8--140 downlink slots versus 82.2--111 measured; independent Bernoulli outcomes overstate runs by omitting residual temporal structure that can readmit an honest UE. The closed-form horizon is 150--219 updates, consistent with long or censored runs. Together, replay and recurrence isolate delivered-service memory as the dominant policy mechanism.

\begin{table}[!tb]
\caption{PF Model Check at $\Delta_r=1$. Win Share and Longest Run Are Measured/Monte Carlo Means over 15 Shards per Target. Horizon Is the Closed-Form Number of Modeled Updates.}
\label{tab:modelcheck}
\centering
\footnotesize
\setlength{\tabcolsep}{3.0pt}
\begin{tabular}{@{}ccccc@{}}
\toprule
Target & \makecell{Win\\share} & \makecell{Mean Absolute\\Error (MAE)} & \makecell{Longest Run\\(Slots)} & Horizon \\
\midrule
UE0 & .950/.995 & .022 & 108.0/140.0 & 219.3 \\
UE1 & .953/.988 & .017 & 82.2/140.0 & 211.8 \\
UE2 & .940/.943 & .007 & 111.0/115.9 & 149.5 \\
UE3 & .940/.951 & .007 & 103.0/120.6 & 154.0 \\
\bottomrule
\end{tabular}
\end{table}

Table~\ref{tab:modelcheck} rules out one favorable target placement: all four targets capture at least 94\% of downlink slots despite different channel-grounded spectral efficiencies and horizons. The agreement supports the analytical mechanism without treating the closed form as an exact stochastic predictor.

\begin{figure}[!tb]
\centering
\includegraphics[width=0.92\columnwidth]{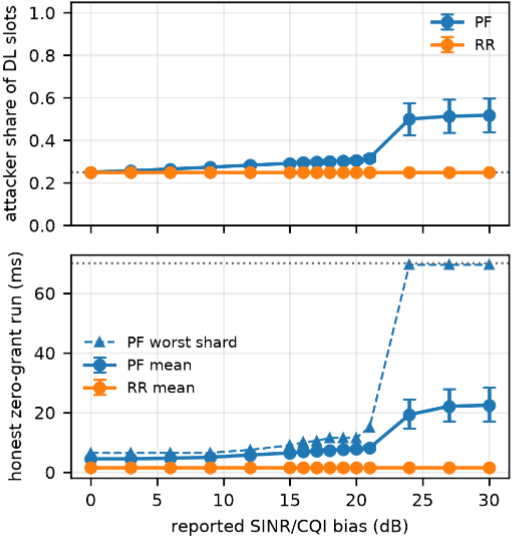}
\caption{Selective CQI/SINR over-reporting on the rank 1 geometry, pooled over 60 shards per point with 95\% bootstrap intervals. At 24--30\,dB, PF share reaches 0.50--0.52 and mean honest runs reach 19--22\,ms; controller exposure ranges from no change under CQI-agnostic SALAD to 0.924--0.958 capture under OLLA. RR remains report-blind.}
\label{fig:cqi-overreport}
\end{figure}

The numerator choice also makes the test conservative with respect to report authority. Equation~\eqref{eq:pf} computes $\mathrm{SE}_i$ from channel-grounded SINR, so neither RI nor CQI manipulation can directly raise the PF numerator. Fig.~\ref{fig:cqi-overreport} shows a weaker denominator trigger under selective CQI/SINR over-reporting on the rank 1 geometry. The pooled PF target share reaches 0.50--0.52 at 24--30\,dB, but the controller ablation is decisive: OLLA reaches 0.924--0.958, SALAD-CQI reaches 0.326--0.345, and vanilla SALAD remains at 0.25 because its MCS path is CQI-agnostic. RR remains at 0.25 with 1.5 ms runs. Failed delivery is therefore the common scheduler mechanism, while exposure depends on which feedback each controller consumes.

\subsection{Per-Slot Mechanism Trace}
\begin{figure}[!tb]
\centering
\includegraphics[width=0.88\columnwidth]{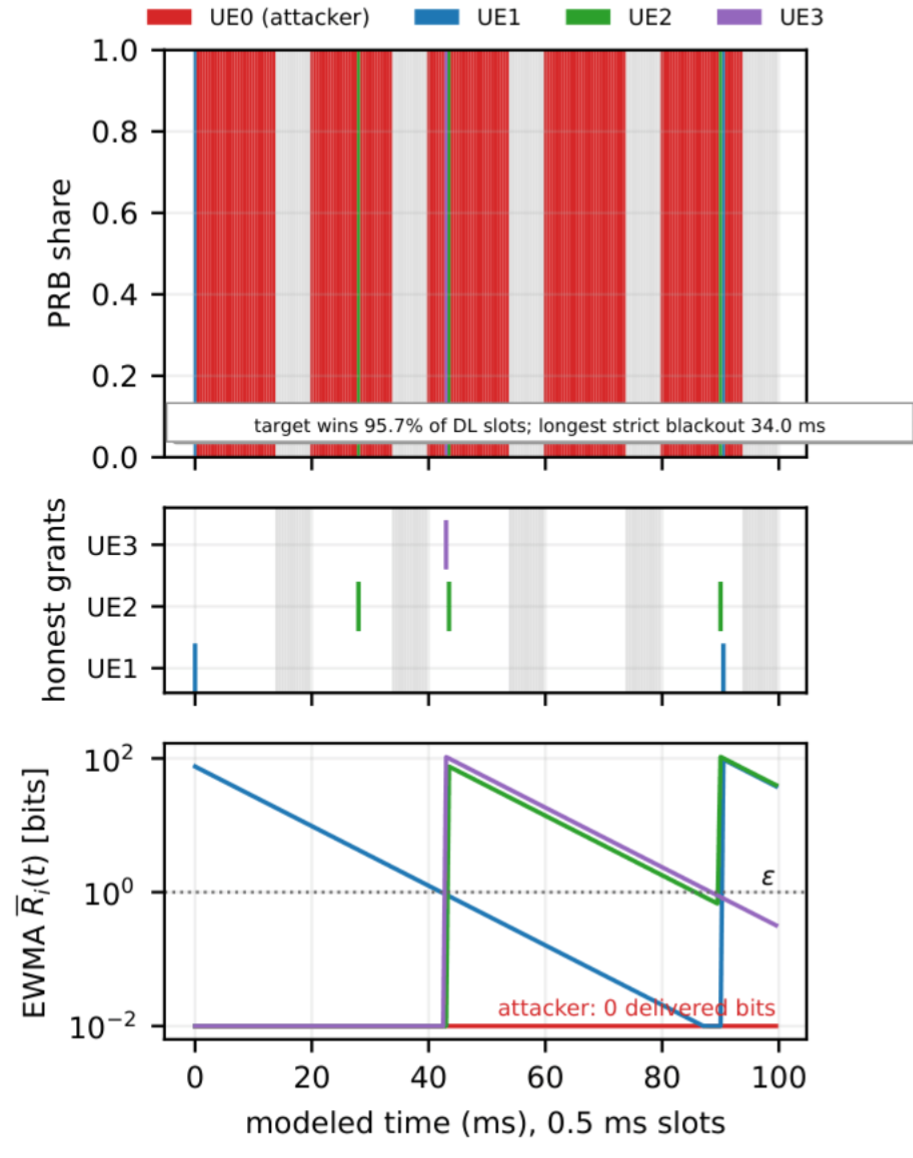}
\caption{Representative selective RI shard ($\Delta_r=1$, OLLA, target UE0, seed 0). The target wins 95.7\% of downlink slots with zero delivered bits; isolated honest grants reset their EWMAs before geometric decay. Gray denotes uplink or special slots; the longest strict blackout is 34.0\,ms.}
\label{fig:ledger}
\end{figure}

Fig.~\ref{fig:ledger} resolves the RI mechanism at individual-slot resolution.
The target receives almost every complete PRB grant, while honest UEs appear as sparse impulses. Its zero delivered bits keep $\overline R_i$ below $\varepsilon$; each honest success raises that UE's EWMA by approximately $\alpha\phi_j$, which then decays until readmission. This is the readmit-and-reset behavior predicted by Eqs.~\eqref{eq:pf}--\eqref{eq:horizon}.

The shard is not a worst case: its target share is 0.957 and longest blackout is 34.0\,ms, below the pooled maxima. It illustrates the mechanism without replacing the 60 shard aggregate and shows how isolated honest grants can improve long-window throughput while contiguous deprivation persists.

\subsection{Observability, Defenses, and Limitations}
The gated-EWMA variants used against bursty traffic can be written with a binary activity gate $g_i[t]$ as $\overline R_i[t+1]=(1-\alpha g_i[t])\overline R_i[t]+\alpha g_i[t]q_i[t]\phi_i[t]$. Under RI inflation, the target is continuously backlogged and receives the full grant, so payload, PRB-threshold, and consecutive-small-grant gates all evaluate to $g_i[t]=1$. There is no idle interval to freeze. Activity gating cannot detect an attacker whose scheduled activity is the attack.

A one-second KPI can average across many 41--56-ms blackouts, whereas the O-DU ledger exposes their continuity. The appropriate control point is therefore upstream of the allocator: compare reported RI with base-station spatial evidence before the report acquires scheduling authority. This follows Zero Trust's principle that authenticated origin does not imply semantic trust and aligns with O-DU-resident dApps and robust MU-MIMO scheduling research \cite{NISTSP800207,ORAN-RR-2024-10-dApps,wu2024r3}. We do not design or evaluate that verifier here.

RR is a causal control, not a proposed replacement for PF. Its report-blind rotation breaks the feedback loop but abandons channel-opportunistic scheduling and still serves an undecodable target. A practical defense should preserve PF for plausible feedback, constrain inconsistent RI, and use a bounded fallback when spatial evidence is uncertain. We identify this control point and its slot-time evidence, not a threshold or actuator.

The scope is intentionally narrow. The allocator is winner-take-all with one virtual PRB pool, $\varepsilon=1$, and $\alpha=0.05$; deployed schedulers can split PRBs and use different memory constants. The harness uses table-based block error rate (BLER), scalar grants, and per-layer BLER aggregation. It excludes coded forward-error correction execution, HARQ process timing, transport-block refitting, radio-frequency injection, fronthaul, and over-the-air realization. The AWGN, TDL-C/D, and CDL-A PHY cells test single UE link adaptation only; the starvation claim comes from modeled-time multi-user RT cells. Four UEs, five seeds per controller, 200 slot shards, and a 70 ms run-length ceiling limit tail resolution. No learning model is trained. These limits make the result a controlled scheduler ablation, not a deployment claim. Future work should repeat it on a timing-faithful inline-accelerated PHY such as Aerial CUDA-Accelerated RAN (ACAR), where graphics processing unit (GPU) decoding headroom and slot deadlines can be measured jointly, and explicit multi-user PHY interference modeling can test whether adversarial CSI induces inter-layer, inter-user, or intra-cell co-channel interference (CCI) in addition to scheduler-mediated starvation \cite{nvidia_aerial_cuda_accelerated_ran,barker2026timessparecharacterizinggpuaccelerated,gadiyar2023inline}.

\balance
\section{Conclusion}
\label{sec:conclusion}
A rank indicator is one integer in a CSI report, but it carries significant scheduling authority. Across all three evaluated link-adaptation controllers and all four target identities, inflating it by one on a channel with spatial headroom drives PF toward 94--95\% target capture and leaves honest UEs without grants for tens of milliseconds. RR prevents PRB capture by construction, although it continues wasting the dishonest UE's allocation. Exact schedule replay and the recurrence model show that the effect is not a report-inflated numerator: failed delivery decays PF's service denominator while the target remains scheduled. The weaker, controller-dependent CQI ablation and the HARQ/PMI specificity controls reinforce that boundary. The evaluated winner-take-all PF allocator thus amplifies RI-induced delivery failure into DoS through honest-user PRB starvation. The next step is causal RI-consistency verification and slot-time enforcement inside the O-DU, followed by validation on a coded, timing-faithful PHY.

\begingroup
\def\BIBdecl{\microtypesetup{expansion=false}}
\bibliographystyle{IEEEtran}
\bibliography{References}
\endgroup

\end{document}